\documentclass[useAMS,usenatbib]{mn2e}
\usepackage{graphicx}
\title[Extended Press-Schechter theory and the density profiles of dark
matter haloes]{Extended Press-Schechter theory and the density
profiles of dark matter haloes}
\author[Nicos Hiotelis]{Nicos Hiotelis\thanks{E-mail:
hiotelis@ipta.demokritos.gr} \thanks{Present address: Roikou
17-19, Neos
Kosmos, Athens, 11743 Greece}\\
1st Experimental Lyceum of Athens, Ipitou 15, Plaka, 10557,
Athens, Greece, E-mail: hiotelis@ipta.demokritos.gr}
\begin{document}

\date{Accepted ............... Received ................; in original form ...........}

\pagerange{\pageref{firstpage}--\pageref{lastpage}} \pubyear{2003}

\maketitle

\label{firstpage}

\begin{abstract}
An inside-out model for the formation of haloes in a hierarchical
clustering scenario is studied. The method combines the picture of
the spherical infall model and a modification of the extended
Press-Schechter theory. The mass accretion rate of a halo is
defined to be the rate of its mass increase due to minor mergers.
The accreted mass is deposited at the outer shells without
changing the density profile of the halo inside its current virial
radius. We applied the method to a flat $\Lambda CDM$ Universe.
The resulting density profiles are compared to analytical models
proposed in the literature, and a very good agreement is found. A
trend is found of the inner density profile becoming steeper for
larger halo mass, that also results from recent N-body
simulations. Additionally, present-day concentrations as well as
their time evolution are derived and it is shown that they
reproduce the results of large cosmological N-body simulations.
\end{abstract}

\begin{keywords}
   cosmology: theory -- dark matter -- galaxies: haloes -- structure -- formation
\end{keywords}
%

\section{Introduction}
Numerical studies  (Quinn, Salmon \& Zurek \cite{quinn}; Frenk et
al. \cite{frenk}; Dubinski \& Galberg \cite{dubinski}; Crone,
Evrard \& Richstone  \cite{crone}; Navarro, Frenk \& White
\cite{navarro}, hereafter NFW; Cole \& Lacey \cite{cole}; Huss,
Jain \& Steinmetz \cite{huss}; Fukushige \& Makino \cite{fuku};
Moore et al. \cite{moore}, hereafter MGQSL; Jing \& Suto
\cite{jing}, hereafter JS: Hernquist \cite{hern90}, hereafter H90,
Kravtsov at al. \cite{kravtsov}, Klypin at al. \cite{klypin}) show
that the density profiles of dark matter haloes are fitted by
models of the form
\begin{equation}
\rho_f(r)=\frac{\rho_c}{(r/r_s)^{\lambda}[1+(r/r_s)^{\mu}]^\nu}
\end{equation}
More specifically NFW proposed a model with $\lambda=1, \mu=1,
\nu=2$, H90 proposed $\lambda =1, \mu=1, \nu=3$, MGQSL
$\lambda=1.5, \mu=1.5, \nu=1$ while  JS proposed
$\lambda=1.5,\mu=1,\nu=1.5$.

The logarithmic slope $\gamma$, of the density profile, defined by
\begin{equation}
\gamma (r)\equiv -
\frac{\mathrm{d}ln\rho_f(r)}{\mathrm{d}r}=\lambda+\mu\nu\frac{(r/r_s)^{\mu}}{1+(r/r_s)^{\mu}}
\end{equation}
 is, in all the above models, a decreasing function of radius.  It is smaller
 than  $2$ near the centre of the system and larger near its virial
 radius.\\

An interesting quantity that characterizes the shape of the
density profile is the concentration. This is defined by
$c=R_\rmn{vir}/r_s$ where $R_\rmn{vir}$ is the virial radius of
the system. In a hierarchical clustering model, the concentration
is a decreasing function of the mass of the system. Thus, smaller
systems, that formed earlier, have higher concentrations than
larger ones. This reflects the high density of the Universe at the
epoch of their formation. The concentration of the resulting
structures is studied in details by large N-body simulations. In
particular, Bullock et al. \cite{bullock}, BKPD hereafter,
constructed a toy-model that describes accurately the time
evolution of the concentration in a way consistent with the
results of their large cosmological simulations.

Although numerical experiments are the most powerful method to
study the formation of structures, the development of analytical
or semi-numerical methods is very important as well since they
help to improve our understanding about the physical processes
during the formation.  Density profiles of equilibrium cold dark
matter haloes are studied by such methods (e.g. Syer \& White
\cite{syer}, Avila-Reese, Firmani \& Hern$\mathrm{\acute{a}}$dez
\cite{avila}, Raig, Gonz$\mathrm{\acute{a}}$lez-Casado \&
Salvador-Sol$\mathrm{\acute{e}}$ \cite{rgcss}, Henriksen \& Widrow
\cite{henriksen}, Nusser \& Sheth \cite{nusser}, Kull \cite{kull},
Lokas \cite{lokas}, Hiotelis \cite{hiot}).

Modifications of the extended Press-Schechter theory (PS) (Press
\& Schechter
 \cite{ps}; Bower \cite{bower}; Bond et al. \cite{bond}, Lacey \&
Cole \cite{lc}) based on the distinction between minor and major
mergers (Manrique \& Salvador-Sol$\mathrm{\acute{e}}$ \cite{mss};
Kitayama \& Suto \cite{kitayama};
Salvador-Sol$\mathrm{\acute{e}}$, Solanes \& Manrique \cite{sssm},
Cohn, Bagla \& White \cite{cohn}) are also used for stydying the
formation of structures. Recently, such a modified PS
approximation was combined with a spherical infall model picture
of formation by Manrique et al. \cite{mrsss}, MRSSS hereafter.
Their results are in good agreement with those of N-body
simulations.

In this paper we use the formalism of MRSSS, with justified
modifications, and the same model parameters as in BKPD. We
compare the characteristics of the resulting structures with those
in N-body results. In Section $2$, we discuss the modified PS
theory and its application to the calculation of the density
profile. In Section 3, the characteristics of the resulting dark
matter haloes are presented.
A discussion is given in Section 4.\\

\section{Extended and Modified Press-Schechter theory}
One of the major goals of the spherical infall model is the PS
approximation. It states that the comoving density of haloes with
mass in the range $M,M+\mathrm{d}M$ at time $t$ is given by the
relation:
\begin{equation}
  N(M,t)\mathrm{d}M=
  \sqrt{\frac{2}{\pi}}\frac{\delta_c(t)}{{\sigma}^2(M)}
  \frac{\rho_{b0}}{M}e^{-\frac{{\delta_c}^2(t)}{2\sigma^2(M)}}
  \mid\frac{\mathrm{d}
 \sigma(M)}{\mathrm{d}M}\mid\mathrm{d}M
 \end{equation}
 where $\sigma(M)$ is the present-day rms mass
 fluctuation on comoving scale containing mass $M$ and is related
  to the power spectrum $P$ by the following relation
 \begin{equation}
 \sigma^2(M) =\frac{2}{{\pi}^2}\int{\hat{W}}^2(kR)P(k) k^2dk
 \end{equation}
 where $\hat{W}$ is the Fourier transform of the window function
 used to smooth the overdensity field. The mass $M$ and the radius $R$ are related
 by the equation
 \begin{equation}
 M=\frac{4}{3}\pi\rho_{b0}R^3=\frac{\Omega_{m0}H^2_0}{2G}R^3
 \end{equation}
 where $\rho_{b0}$ is the present-day value of the density of the
 unperturbed Universe, $\Omega_{m0}$ is the present-day value of
 the density parameter (defined at any scale factor
 $a$ by the relation $\Omega_m(a)=8\pi G \rho_b(a)/(3H^2(a)) )$
  and  $H_0$ is the present-day value of Hubble's
 constant, $H$.

 The only time dependent term of Eq. 3 is $\delta_c(t)$ that is the
 linear extrapolation up to the present epoch of the primordial
 density that collapses at $t$. It is calculated using the following
 arguments: In a model universe with cosmological constant
 $\Lambda$, the radius $r$ of a sphere  having  initial  overdensity
 $\Delta_i$, evolves according to the equation
 \begin{equation}
 \frac{\mathrm{d}s}{\mathrm{d}t}=H_ig^{\frac{1}{2}}(s)
 \end{equation}
 where $s\equiv r/r_i$ and $r_i$ is the initial radius.
 $H_i$ is the value of the Hubble's constant
 at the initial time  $t_i$ and $g$ is given by the relation
 \begin{equation}
 g(s)=\Omega_{m,i}(1+\Delta_i)(s^{-1}-1)+\Omega_{\Lambda,i}(s^2-1)+1-\frac{2}{3}f_i\Delta_i
 \end{equation}
and $\Omega_{\Lambda,i}$ is the initial values of
 the quantity $\Omega_{\Lambda}(a)=\Lambda/(3H^2(a))$.
Eq.7 is derived under the assumption that the initial velocity
$v_i$ of the shell is $v_i=H_ir_i-v_{pec,i}$ where the initial
peculiar velocity,$v_{pec,i}$, is given according to the linear
theory by the relation $v_{pec,i}=\frac{1}{3}H_ir_if_i\Delta_i$
~(Peebles \cite{peebles}). A very good approximation of the factor
$f_i$ is
$f_i={\Omega_{m,i}}^{0.6}+\frac{1}{70}[1-\Omega_{m,i}(1+\Omega_{m,i})]$
(Lahav et al. \cite{lahav}). The radius of the maximum expansion
is $r_{ta}=s_{ta}r_i$, where $s_{ta}$ is the root of the equation
$g(s)=0$  that corresponds to zero velocity
($\mathrm{d}s/\mathrm{d}t=0$). The sphere reaches its turn-around
radius at time
\begin{equation}
t_{ta}=\frac{1}{H_i}\int_0^{s_{ta}}g^{-\frac{1}{2}}(s)\mathrm{d}s
\end{equation}
and then collapses at time $t_{c}=2t_{ta}$.

 The scale factor $a$
of the Universe obeys the equation:
\begin{equation}
\frac{\mathrm{d} a}{\mathrm{d} t}=H_0X^{\frac{1}{2}}(a)
\end{equation}
 where
 \begin{equation}
X(a)=1+\Omega_{m,0}(a^{-1}-1)+\Omega_{\Lambda,0}(a^2-1)
\end{equation}
 and the subscript $0$ denotes the present-day values
of the parameters. However, the time and the scale factor $a$ are
related by the equation
\begin{equation}
t=\frac{1}{H_0}\int_0^{a }X^{-\frac{1}{2}}(u)\mathrm{d}u
\end{equation}
Setting $t=t_c$ in the  above equation and solving for $a$ one
finds the scale factor $a_c$ at the epoch of collapse. If we call
$\delta_{i,c}(t)$ the initial overdensity required for the
spherical region to collapse at that time $t$ and take into
account the linear theory for the evolution of the matter density
contrast $\delta=\delta\rho/\rho$, we have
\begin{equation}
\delta\propto\frac{1}{H_0^2}\frac{X^{1/2}(a)}{a}\int_0^{a}X^{-3/2}(u)\mathrm{d}u=D(a),
\end{equation}
(Peebles 1980), then $\delta_c(t)$ is given
\begin{equation}
\delta_c(t)=\delta_{i,c}(t)\frac{D(t_0)}{D(t_i)},
\end{equation}
where $t_0$ denotes the present epoch.\\
Usually $\delta_c$ is written in the form
\begin{equation}
\delta_c(t)=\delta_{i,c}(t)\frac{D(t)}{D(t_i)}\frac{D(t_0)}{D(t)}=\delta_\rmn{crit}(t)\frac{D(t_0)}{D(t)}
\end{equation}
where $\delta_\rmn{crit}(t)$ is the linear extrapolation of the
initial overdensity up to the time $t$ of its collapse. In an
Einstein-de Sitter universe ($\Omega_m=1,\Omega_{\Lambda}=0$) this
value is independent on the time of collapse and is
$\delta_\rmn{crit}\approx 1.686$. In other cosmologies it has  a
weak dependence on the time of collapse (e.g. Eke, Cole \& Frenk
\cite{eke}). In a flat universe it can be approximated by the
formula $\delta_\rmn{crit}(t)\approx
1.686\Omega^{0.0055}_{m,0}(t)$.

The PS mass function agrees relatively well with the results of
N-body simulations (e.g. Efstathiou, Frenk \& White \cite{efst1},
Efstathiou \& Rees \cite{efst2}; White, Efstathiou \& Frenk
\cite{wef}, Lacey \& Cole \cite{lc94}; Gelb \& Bertschinger
\cite{gelb}; Bond \& Myers \cite{bondm}) while it deviates in
detail at both the high and low masses. Recent improvements (Sheth
\& Tormen \cite{sheth}; Sheth, Mo \& Tormen \cite{shethetal}, see
also Jenkins et al.\cite{jenkins}) allow a better approximation
involving some more parameters. The application of the above
approximation to the model studied in this paper is a subject of
future research.

Lacey \& Cole \cite{lc} extended the PS theory using the idea of a
Brownian random walk, and were able to calculate analytically
tractable expressions for the mass function, merger rates, and
other properties. They show that the instantaneous transition rate
at $t$ from haloes with mass $M$ to haloes with mass between
$M',M'+\mathrm{d}M'$ is given by
\begin{eqnarray}
  r(M\rightarrow
  M',t)\mathrm{d}M'=\nonumber\\
  \left(\frac{2}{\pi}\right)^{1/2}\frac{\mathrm{d}\delta_c(t)}{\mathrm{d}t}
  \frac{1}{\sigma^2(M')}\frac{\mathrm{d}\sigma(M')}{\mathrm{d}M'}
  \left[1-\frac{\sigma^2(M')}{\sigma^2(M)}\right]^{-3/2}\nonumber\\
  \times\exp\left[-\frac{\delta^2_c(t)}{2}
  \left(\frac{1}{\sigma^2(M')}-\frac{1}{\sigma^2(M)}\right)\right]\mathrm{d}M'
  \end{eqnarray}
  This provides the fraction of the total number of haloes with
  mass $M$ at $t$, which give rise per unit time  to haloes with mass in
  the range $M', M'+\mathrm{d}M'$ through instantaneous mergers of any
  amplitude.

  An interesting modification of the extended PS theory is the distinction
  between minor and major mergers
(Manrique \& Salvador-Sol$\mathrm{\acute{e}}$ \cite{mss};
Kitayama \& Suto \cite{kitayama};
  Salvador-Sol$\mathrm{\acute{e}}$ et al. \cite{sssm}; Percival,
  Miller \& Peacock \cite{percival}, Cohn et al. \cite{cohn}).

  Mergers that produce a fractional increase below a given threshold $\Delta_m$
  are regarded as minor. This kind of mergers corresponds to an
  accretion.  Consequently, the  rate
  at which haloes increase their mass due to minor mergers is the
  instantaneous mass accretion rate and is given by the relation
  \begin{equation}
     r^a _{m}(M,t)=\int_M^{M(1+\Delta m)}(M'-M)r(M\rightarrow
   M',t)\mathrm{d}M'
   \end{equation}
   Thus the rate of the increase of halo's mass due to the accretion is
   \begin{equation}
   \frac{\mathrm{d}M(t)}{\mathrm{d}t}=r^a_{m}[M(t),t]
   \end{equation}
Before proceeding further with the model, it is useful to
   discus briefly the cosmological considerations about the virial radius of
   a spherical system.
   Let $\Delta_\rmn{vir}(a)$ be the ratio of the overdensity
   of a sphere, that has collapsed and virialized at scale factor
   $a$, to the background density. This can be expressed by the form:
   \begin{equation}
   \Delta_\rmn{vir}(a)=\frac{\rho(a)}{\rho_b(a)}=\frac{1}{s^3_{ta}c^3_f}\left(\frac{a}{a_i}\right)^3(1+\Delta_i)
   \end{equation}
   where $c_f$ is the collapse factor of the sphere defined as the ratio of its
   final radius to its turnaround radius. Lahav et al. \cite{lahav} applied the virial
   theorem to the virialized final sphere assuming a flat overdensity
   and found the collapse factor to be
   $c_f\approx (1-n/2)/(2-n/2)$ where $n=(\Lambda
   r^3_{ta})/(3GM)$.
   For an Einstein-de Sitter Universe $\Delta_\rmn{vir}(a)\approx
   18{\pi}^2$ at any time. For flat models with cosmological
   constant,
   significantly good analytical approximations of $\Delta_\rmn{vir}$ exist.
   Bryan \& Norman \cite{bryan} proposed for $\Delta_\rmn{vir}$
 the following approximation
   \begin{equation}
   \Delta_\rmn{vir}(a)\approx(18{\pi}^2-82x-39x^2)/\Omega_m(a)
   \end{equation}
   where $x\equiv 1-\Omega_m(a)$.

   MRSSS considered the following picture of the formation of a
   halo:
   At time $t_i$ an halo of virial mass  $M_i$ and virial radius $R_i$ is
   formed and at later times it accretes mass according to the
   Eq.(17). Assuming that the accreted mass is deposited in an outer spherical shell
   without changing the density profile inside its current radius,
   then
   \begin{equation}
   M_\rmn{vir}(t)-M_i=\int_{R_i}^{R_\rmn{vir}(t)}4\pi r^2\rho(r)\mathrm{d}r
   \end{equation}
   The current radius $R_\rmn{vir}$ contains a
   mass with mean density $\Delta_\rmn{vir}(a)$ times the mean density of the Universe
   $\rho_b(t)$. Therefore,
   \begin{equation}
   R_\rmn{vir}(t)=\left[\frac{3M_\rmn{vir}(t)}{4\pi
   \Delta_\rmn{vir}(t)\rho_b(t)}\right]^{1/3}.
   \end{equation}
   Differentiating with respect to $t$
   \begin{equation}
   \rho(t)=\Delta_\rmn{vir}(t)\rho_b(t)\left[1-\frac{M_\rmn{vir}(t)}{r^a_{m}[M_\rmn{vir}(t),t]}
   \frac{\mathrm{d}[\ln[\rho_b(t)\Delta_\rmn{vir}(t)]]}{\mathrm{d}t}\right]^{-1}
   \end{equation}
   Since one of the goals of this paper is the comparison of our results with
   the results of the N-body simulations of BKPD,
   we have used for $\Delta_\rmn{vir}(t)$ the same approximation as
   BKPD did -that is Eq.19- and not the constant value of 200 that MRSSS used.
   It is convenient to express Eq.22 in terms of the scale factor $a$ instead of
   the time $t$. Thus, Eq.22 becomes:\
   \begin{eqnarray}
   \rho(a)=\frac{3H^ 2(a)\Omega_m(a)}{8\pi
   G}\Delta_\rmn{vir}(a)\times \nonumber\\
   \left[1+\frac{M_\rmn{vir}(a)}{r^{acc}_{m}[M_\rmn{vir}(a),a]}
   \left(\frac{3}{a}-\frac{\mathrm{d}\ln[\Delta_\rmn{vir}(a)]}{\mathrm{d}a}\right)\right]^{-1}
   \end{eqnarray}
   where we used $\rho_b(a)a^3=const.$ and
   \begin{equation}
   r^{acc}_{m}[M_\rmn{vir}(a),a]\equiv\frac{\mathrm{d}M_\rmn{vir}}{\mathrm{d}a}=H_0^{-1}X^{-1/2}(a)r^{a}_{m}[M_\rmn{vir}(t),t].
   \end{equation}
   Integrating Eq.17 and using Eqs. $21$ and $23$, we obtain the growth of virial mass
   and virial radius and, in a parametric form, the density
   profile of haloes.

\section{Density profiles of dark matter haloes}

The results described in this section are derived for a flat
universe with $\Omega_{m,0}=0.3$ and $\Omega_{\Lambda,0}=0.7$. We
used two forms of power spectrum. The first one -named spect1- is
the one proposed by Efstathiou, Bond \& White \cite{ebw}. It is
based on the results of the COBE DMR experiment and is given by
the relation:
\begin{equation}
P_{spect1}(k) =\frac{Bk}{[1+[ak+(bk)^{3/2}+(ck)^2]^{\nu}]^{2/\nu}}
\end{equation}
where $a=(6.4/\Gamma)h^{-1} \rmn{Mpc},~ b=(3.0/\Gamma)h^{-1}
\rmn{Mpc},~ c=(1.7/\Gamma)h^{-1} \rmn{Mpc}$ and $\nu=1.13$.
Low-density Cold Dark Matter (CDM) models in a spatially flat
Universe (i.e. $\Lambda>0$) are described for
$\Gamma=\Omega_{m,0}h$.

The second spectrum -named spect2- is the one proposed by Smith et
al. \cite{smith} and is given by:
\begin{equation}
P_{spect2}(k)
=\frac{Ak^n}{[1+a_1k^{1/2}+a_2k+a_3k^{3/2}+a_4k^2]^b}
\end{equation}
The values for the parameters are: $n=1,~ a_1=-1.5598,~
a_2=47.986, ~a_3=117.77,~ a_4=321.92$ and $b=1.8606$.

\begin{figure}
\includegraphics{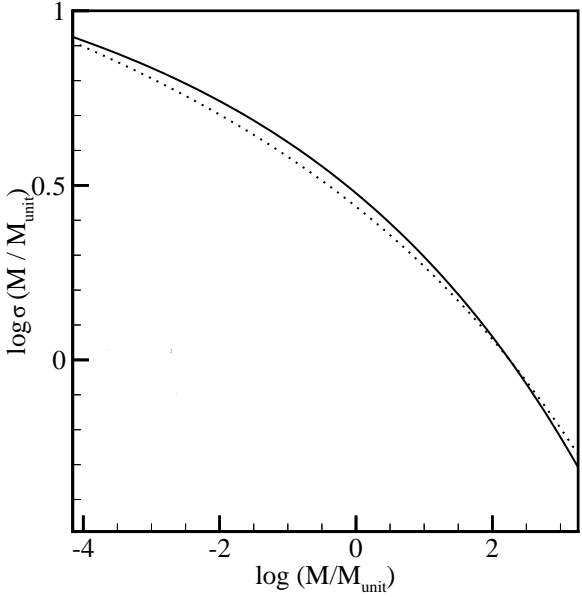}
 \caption{rms mass fluctuation as a function of mass. Solid line: for spect1,
  dotted line: for spect2}
\label{fig1}
\end{figure}

We used the top-hat window function that has a Fourier transform
given by:
\begin{equation}
\hat{W}(kR)=\frac{3(\sin(kR)-kR\cos(kR))}{(kR)^3}
\end{equation}
The constants of proportionality $A$ and $B$ are found using the
procedure of normalization for $\sigma_8\equiv
\sigma(R=8h^{-1}\rmn{Mpc})=1.$ In Fig.\ref{fig1}, the resulting
rms fluctuations for both spectra are shown. It must be noted that
we use a system of units with
$M_\rmn{unit}=10^{12}h^{-1}\rmn{M_{\sun}},~
 R_\rmn{unit}=h^{-1}\rmn{Mpc} $ and $t_\rmn{unit}=1.515\times10^{7}h^{-1}\rmn{years}$.
  In this system of units
$H_0/H_{\rmn{unit}}=1.5276$.

\begin{figure}
\includegraphics{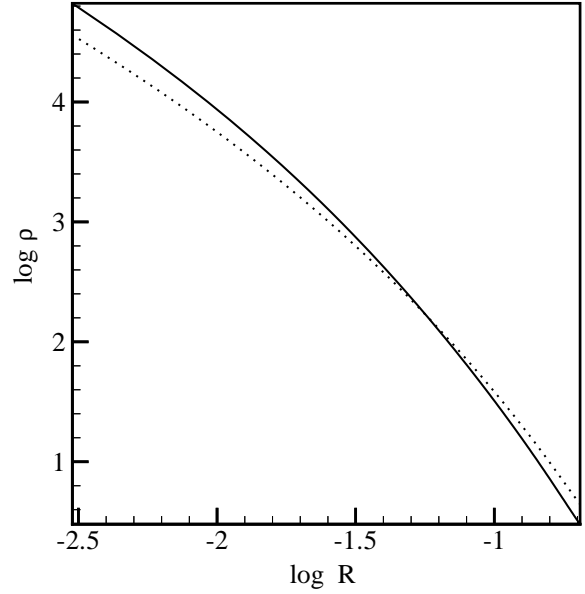}
 \caption{For spect2: Density profiles for two haloes having the
 same mass $10^{12}h^{-1}\rmn{M}_{\sun}$. Solid line: $\Delta_m
=0.21$.  Dotted line: $\Delta_m=0.5$}
 \label{fig2}
\end{figure}

\subsection{Present day structures}
In the approximation used in this paper, for given models of the
Universe and power spectrum there is only one free parameter, that
is the value of the threshold $\Delta_m$ (see Eq.16). We found
that the resulting density profiles are sensitive to the value of
$\Delta_m$. As an example, the density profiles of two systems
with the same present-day mass $10^{12}h^{-1}\rmn{M}_{\sun}$ and
different values of $\Delta_m$ are plotted in Fig.\ref{fig2}. Both
density profiles are derived from spect2. The solid line -shown in
Fig.\ref{fig2}- corresponds to the system derived for $\Delta_m
=0.21$ while the dotted line to the system for $\Delta_m=0.5$. The
density profile for smaller $\Delta_m$ is steeper at both the
inner and the outer regions. Additionally, for different
$\Delta_m$ the concentrations of the haloes are different too (a
detailed description of the way the concentration is calculated is
given below). The system that results for $\Delta_m =0.21$ has
$c_\rmn{vir}=15.2$, while the one for $\Delta_m=0.5$ has
$c_\rmn{vir}=8.7$. In order to calculate the density profiles
(that will be presented below), we used as a basic criterion the
concentrations of the present-day structures. In fact, we have
chosen the values of $\Delta_m=0.23$ and $\Delta_m=0.21$ for
spect1 and spect2 respectively, because the concentrations
resulting from these values are close to the results of the
toy-model of BKPD. This model is constructed by BKPD to reproduce
the results of their N-body simulations and it is able to give the
concentration $c_\rmn{vir}$ of a virial mass $M_\rmn{vir}$ at any
scale factor $a$. First, the scale factor $a_c$ at the epoch of
collapse is calculated,  solving the following equation
\begin{equation}
\sigma[M_{*}(a_c)]=\sigma[FM_\rmn{vir}(a)]
\end{equation}
 where $F=0.01$ and $M_{*}$ is the typical collapsing mass.
 Then,  the  concentration is calculated using the formula
 \begin{equation}
 c_\rmn{vir,BKPD}[M_\rmn{vir}(a),a]=K\frac{a}{a_c}
 \end{equation}
 where $K=4$.
 We recall that the typical collapsing mass at scale factor $a$ satisfies
 $\sigma[M_*(a)]=1.686D(1)/D(a)$.

 It is obvious that the above defined concentration depends only on the cosmology
 and the power spectrum used.
 Thus, for given cosmology and halo mass, the concentration  $c_\rmn{vir,BKPD}$
  is known \textit{a priori} without
 taking into consideration any particular form of halo growth.
  We applied this toy-model to find $c_\rmn{vir,BKPD}$ for the present-day structures.

\begin{figure}
\includegraphics{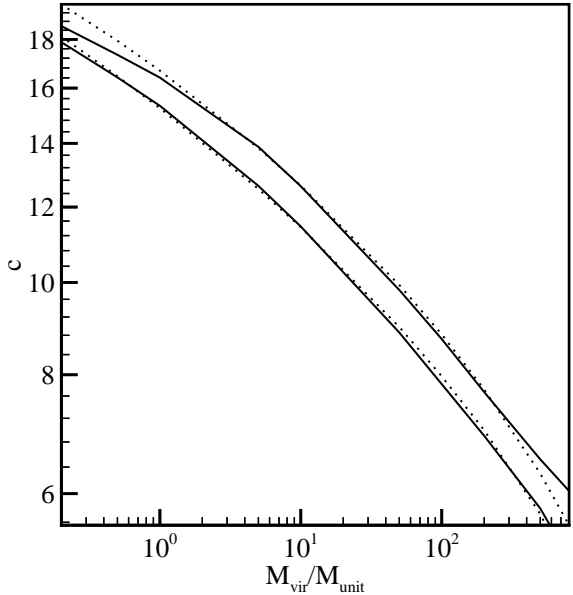}
 \caption{Concentration as a function of present-day virial mass.
 From the top, the first pair of curves are for spect1 and the
 second for spect2. Solid curves: our results. Dotted curves: BKPD
 toy-model results.}
\label{fig3}
\end{figure}

 Another way to calculate the concentration is by using
  $c_\rmn{vir}=R_\rmn{vir}/r_2$ where $r_2$ is the radius
 where the logarithmic slope of the density profile equals 2. This radius is found
 by the following procedure:
 First, the resulting density profiles are fitted by the general
 formula of Eq.1. This is done by minimizing the sum
 \begin{equation}
 S=\sum[\rho(r)-\rho_f(r)]^2.
 \end{equation}
 where $\rho_c$, $r_s$, $\lambda$, $\mu$ and $\nu$ are fitting
 parameters. The minimization is performed using the unconstrained subroutine
 ZXMWD of IMSL mathematical library.
  Then, $r_2$ is found by applying the following formula for $n=2$
 \begin{equation}
 r_n=\left[\frac{n-\lambda}{\lambda+\mu\nu-n}\right]^{1/\mu}r_s
 \end{equation}
 This formula gives the radius $r_n$ at which the logarithmic slope equals
 to $n$.
 According to the model presented in this paper, haloes grow inside-out.
 Thus, the value of  $c_\rmn{vir}$ represents the way of halo
 growth.  In Fig. \ref{fig3}, the concentration is plotted as a function of the
 present-day virial mass. From the top of the
 figure, the first pair of curves (solid and dotted) correspond to
 spect1 and the second pair to spect2.
 Solid curves show our results while dotted curves depict the
 results of the toy-model of BKPD.
 A  very good agreement between the values of the concentration is shown.
 In particular,
concentrations resulting from spect2 are in agreement with those
obtained for
 the model of BKPD for the whole range of mass presented. On the other
 hand, small differences appear for very small and very large masses in the case of
 spect1.

\begin{figure}
\includegraphics{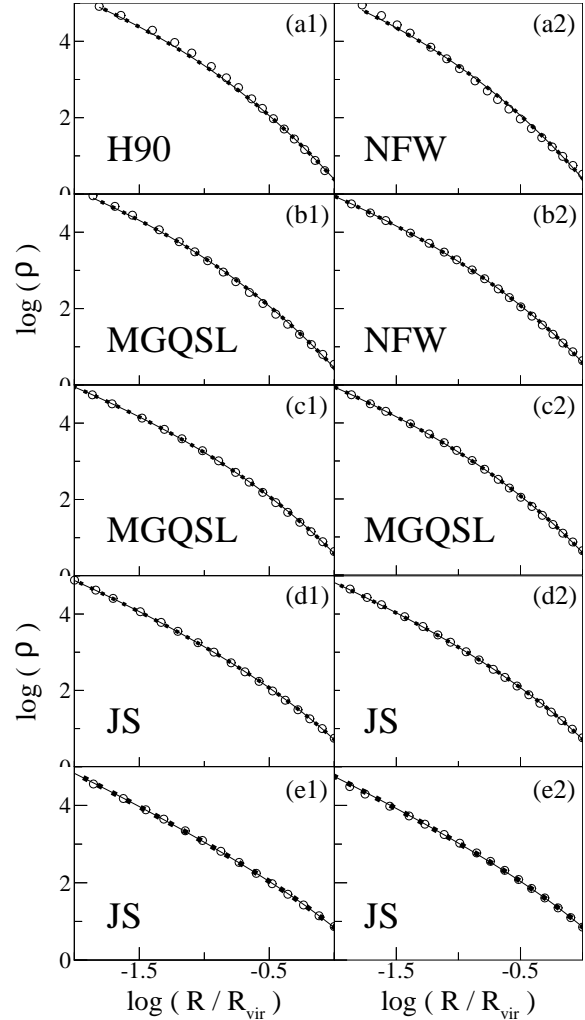}
 \caption{Density profiles as a function of radius.
Solid curves: resulting density profiles. Dotted curves: fits of
the resulting density profiles using the formula of Eq.1. Circles:
best bit to our results using models proposed in the literature
(H90, NFW, MGQSL, JS). Left-hand side: spect1. Right-hand side:
spect2} \label{fig4}
\end{figure}

In Fig. \ref{fig4} we present the density profiles of the
resulting structures with present-day masses in the range of
$0.2\times10^{11}h^{-1}\rmn{M}_{\sun}$ to
$8\times10^{14}h^{-1}\rmn{M}_{\sun}$. The left-hand side figures
(a1, b1, c1, d1, e1) have been produced using spect1, while the
right-hand ones using spect2. Figures (a1) and (a2) correspond to
mass $0.2\times10^{11}h^{-1}\rmn{M}_{\sun}$, (b1) and (b2) have
mass $10^{12}h^{-1}\rmn{M}_{\sun}$, (c1) and (c2) to mass
$10^{13}h^{-1}\rmn{M}_{\sun}$, (d1) and (d2) to mass
$10^{14}h^{-1}\rmn{M}_{\sun}$ and (e1) and (e2) correspond to mass
$8\times10^{14}h^{-1}\rmn{M}_{\sun}$.
 Solid lines represent the resulting density profiles while
 dotted lines are the fits using the general formula of Eq.1.
 It is shown that the fits using the general formula of Eq.1 are exact.
We also fit every halo density profile using the analytical models
that have been proposed in the literature (H90, NFW, MGQSL, JS)
and are described in Section 1. The best fit of these models to
our resulting profiles is shown in Fig. \ref{fig4} (circles). This
best fit is found by the minimizing procedure described above, for
$\lambda, \mu $ and $\nu$ constants and equal to the proposed
values, while $\rho_c$ and $r_s$ are the only fitting parameters.
Best fit for the resulting density profile in (a1) is the H90
model, in (a2) and (b2) the NFW model, in (b1), (c1) and (c2) the
MGQSL model and in (d1), (d2), (e1) and (e2) the JS model.
Additionally, haloes of different mass are fitted well by
different analytical models. This is due to the different inner
and outer slopes of the density profiles. Inner slope, (defined as
that at radial distance $r=10^{-2}R_\rmn{vir}$), is an increasing
function of the virial mass of the halo. For example, in the case
of spect2 the inner slope varies from $1.43$ for
$M=10^{12}h^{-1}\rmn{M}_{\sun}$ to $1.65$ for $M=8 \times
10^{14}h^{-1}\rmn{M}_{\sun}$. Additionally, outer slope -at
$r=R_\rmn{vir}$- is a decreasing function of the virial mass and
it varies from $3.67$ to $2.64$ for the above range of masses.
 \begin{figure}
\includegraphics{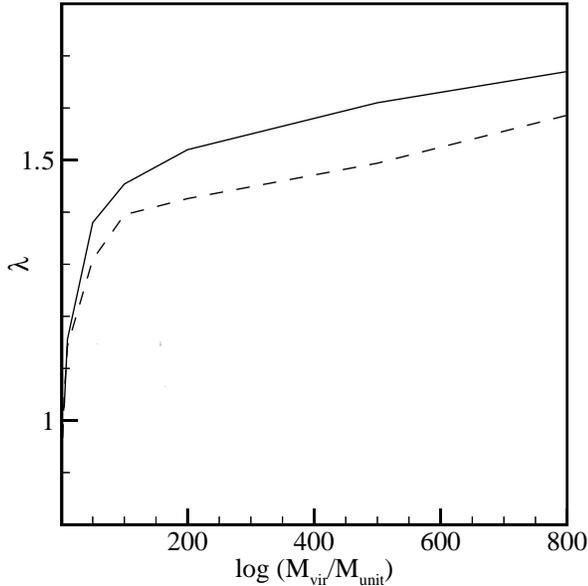}
 \caption{Exponent $\lambda$ as a function of present-day
virial mass for both power spectra. It is shown that $\lambda$ is
an increasing function of the virial mass. Solid curve: spect1.
Dashed curve: spect2} \label{fig5}
\end{figure}

Although density profiles resulting in simulations seem to be
similar, systematic trends that relate them with the power
spectrum have been reported. For example, Subramanian, Cen \&
Ostriker \cite{subra} found in the results of their N-body
simulations the following: for power spectra of the form
$P(k)\propto k^n$ the density profiles have steeper cores for
larger $n$. Therefore, a dependence of the density profile on the
power spectrum is expected. This dependence is shown in our
results comparing the profiles of haloes with the same present day
mass. It should be noted that the method studied in this
manuscript is applicable for the era of slow accretion when the
infalling matter is in the form of small haloes that have mass
less than $\Delta m$ times the mass of the parent halo. This kind
of accretion occurs at the late stages of formation and thus
determines the profile of the outer regions of the halo under
study. However, the values of the inner slopes may be
questionable. Real haloes have followed different mass growth
histories and thus their properties show a significant scatter
about a mean value. Unfortunately, the method studied in this
manuscript results one profile for a halo of given mass. Thus, its
purpose is just to approximate the mean density profile of a large
number of mass growth histories. Since the mass growth history
resulting from the method is in good agreement with the mean
growth history resulting from N-body simulation -as it will be
shown below- then the values of the inner slopes could be close to
the ones of N-body simulations. A Monte Carlo analytic approach
based on the construction of a large number of mass accretion
histories is under study. This study could answer to some of the
above problems.

In Fig. \ref{fig5} the exponent $\lambda$ is plotted, that
  gives the asymptotic slope at $R\rightarrow 0$,
  derived by the general fit as a function of present-day
virial mass for both power spectra. It is shown that the exponent
$\lambda$ is an increasing function of virial mass. This trend of
the inner density profile is also found in the results of recent
N-body simulations (Ricotti \cite{ricotti}).

 \subsection{Time evolution}
\begin{figure*}
\includegraphics{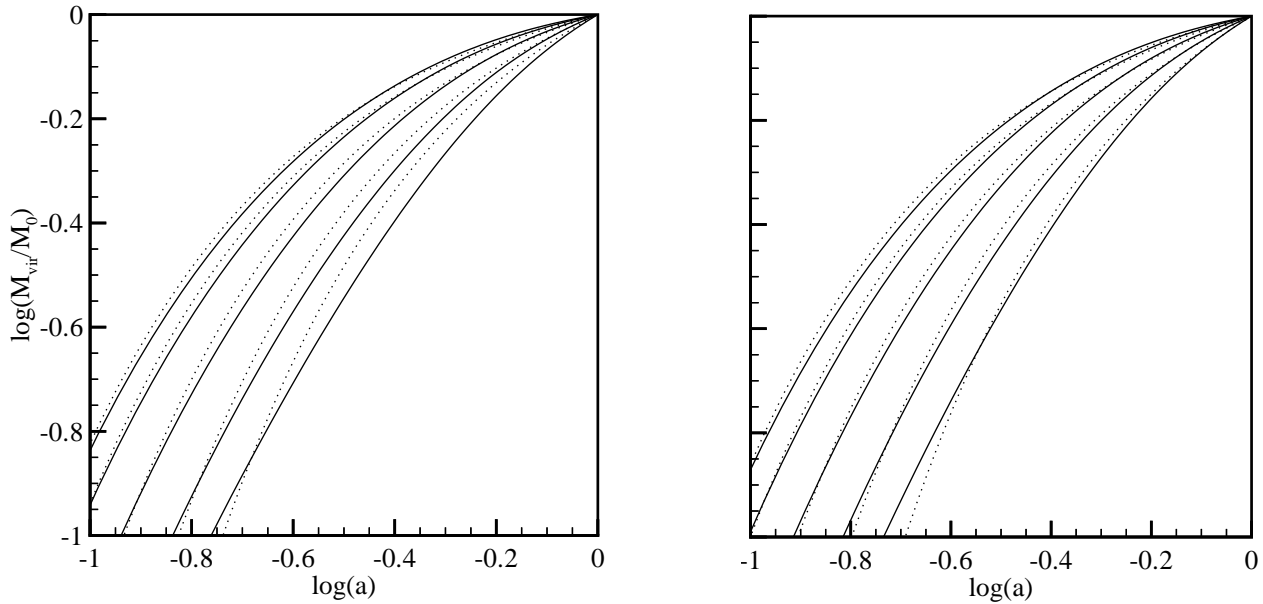}
 \caption{Mass growth curves as a function of scale factor $a$.
 Left panel, right panel: spect1, spect2 respectively.
 From top to the bottom the curves correspond to masses
 $0.2\times10^{11}h^{-1}\rmn{M}_{\sun}$,
 $10^{12}h^{-1}\rmn{M}_{\sun}$,
$ 10^{13}h^{-1}\rmn{M}_{\sun}$, $10^{14}h^{-1}\rmn{M}_{\sun}$
 and $8\times10^{14}h^{-1}\rmn{M}_{\sun}$ respectively.}
\label{fig6}
\end{figure*}

In Fig. \ref{fig6} we plot mass growth curves. The curves show
$M_\rmn{vir}(a)$ as a function of
 $a$ in a logarithmic slope. The solid lines show our resulting structures
 and the dotted lines show the  mass growth curves
 of the model proposed by Wechsler et al. \cite{we}.
  The curves of the left panel correspond to spect1 while those of the right
 panel to spect2. From the top to the bottom, the curves correspond to masses
 $2\times 10^{11}h^{-1}\rmn{M}_{\sun}$, $10^{12}h^{-1}\rmn{M}_{\sun}$,
 $ 10^{13}h^{-1}\rmn{M}_{\sun}$,
 $10^{14}h^{-1}\rmn{M}_{\sun}$ and $8\times 10^{14}h^{-1}\rmn{M}_{\sun}$ respectively.
 It is obvious
 that massive haloes show substantial increase of their mass up to late times while the
 growth curves of less massive haloes tend to flatten out earlier. This behaviour of mass
  growth curves characterizes the hierarchical clustering
 scenario where small haloes are formed earlier than more massive
 ones. Additionally, it helps to define the term "formation time" by
 a measurable way. Wechsler et al. \cite{we} define as formation
 scale factor $\tilde{a}_c$ the scale factor when the logarithmic slope of mass growth,
 ($\mathrm{d}lnM(a)/\mathrm{d}lna$), falls
 below some specified value, $S$. They use the value $S=2$. It should be noted
 that this definition of formation scale factor differs from
 $a_c$, defined by BKPD, since $a_c$ is the value of the scale
 factor at the epoch the typical collapsing mass is $F$ times the
 virial mass of the halo. We found that the values of $\tilde{a}_c$ and $a_c$
 for $F=0.01$ and $S=2$ are different. This is also noticed in
 Wechsler et al. \cite{we} since they state that $\tilde{a}_c$ ~and
 $a_c$ have similar values for $S=2$ but for $F=0.015$. However, the use of the
 value $F=0.015$ in the toy-model of BKPD changes the resulting
 concentrations and so our basic criterion for the choice of the
 threshold $\Delta_m$ is not satisfied. Therefore, it is preferable to
 choose a different value of $S$ for the definition of
 $\tilde{a}_c$, that of $S=1.5$.
 In Fig.\ref{fig6}, the dotted lines  show the  mass growth curves
 of the model proposed by Wechsler et al. \cite{we}. In this model
 the mass growth is calculated using the relation:
 \begin{equation}
 M_\rmn{vir}(a)=M_{vir,0}\exp[-\tilde{a}_cS(1/a-1)]
 \end{equation}
 where $M_{vir,0}$ is the present-day virial mass and the formation scale factor
 $\tilde{a}_c$ is defined by the condition  $\mathrm{d}lnM(a)/\mathrm{d}lna =S$ with
 $S=1.5$. In Fig. \ref{fig6},
  a very satisfactory agreement is shown, particularly for the less massive haloes.
  \begin{figure*}
\includegraphics{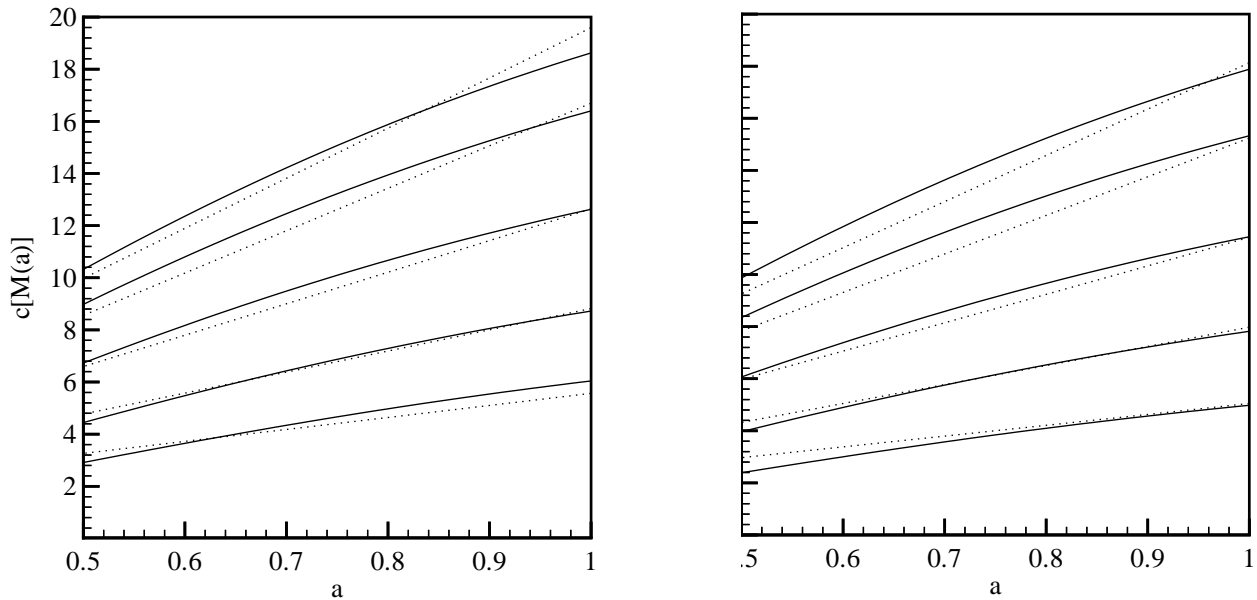}
 \caption{Concentration as a function of the scale factor $a$. Left panel: spect1.
 Right panel: spect2. Solid lines: our results.
 Dotted lines: the results of toy-model of BKPD. From top to the bottom
 the lines correspond to masses $0.2\times 10^{11}h^{-1}\rmn{M}_{\sun}$,
  $10^{12}h^{-1}\rmn{M}_{\sun}$,
 $ 10^{13}h^{-1}\rmn{M}_{\sun}$,
 $10^{14}h^{-1}\rmn{M}_{\sun}$ and $8\times 10^{14}h^{-1}\rmn{M}_{\sun}$ respectively.}
\label{fig7}
\end{figure*}

We have to note that our model haloes grow inside-out. Therefore,
in early enough times -when the slope
 of the density is smaller that $2$ all the way from the centre up
 to the current radius- it is meaningless to define $c_{vir}$.
 Once the building of the halo has proceeded beyond the point with slope 2,
  the evolution
 of $c_\rmn{vir}$ is due to the growth of the virial radius and is given by
 \begin{equation}
 c_\rmn{vir}[M(a),a]=c_\rmn{vir}(M_0)\frac{R_\rmn{vir}[M(a)]}{R_\rmn{vir}(M_0)},
 \end{equation}
 where $c_\rmn{vir}(M_0)$ denotes the present-day concentration and
 $R_\rmn{vir}[M(a)]$ and $R_\rmn{vir}(M_0)$ are the values of the
  virial radius at scale factor $a$ and at the present-day respectively.
   In Fig. \ref{fig7} the
 time evolution of concentrations is plotted. Solid lines
 describe $c_\rmn{vir}$ while dotted lines are $c_\rmn{vir,BKPD}$.
 More massive haloes have lower concentrations that
 evolve slower, while the concentrations of less massive haloes are
 higher and evolve more
 rapidly.

\section{Conclusions}
Since the formation of structures in a hierarchical clustering
scenario is a complicated process, any attempt for the
construction of analytical models requires a number of crucial
assumptions.

The model studied in this paper was proposed by MRSSS and assumes
that
\begin{enumerate}
  \item The rate of mass accretion is defined by the rate of minor
  mergers
  \item Haloes grow inside-out. The accreted mass is deposited at
  the outer shells without changing the density profile of the
  halo inside its current virial radius
  \end{enumerate}
  The first assumption indicates that structures presented
  in this paper formed by a gentle accretion of mass.
   The physical process implied by
  the second assumption is that the infalling matter does not
  penetrate the current virial radius. This process requires an
  amount of non-radial motion. This amount has to be large enough
  so that the pericenter of the accreted mass is larger than the current
 virial radius. It should be noted that a density profile that results
  from a radial collapse has inner slope steeper than $2$.
  It is the presence of non-radial motion during the collapse that
  leads  to inner slopes shallower than $2.$ (e.g. Nusser \cite{nus},
  Hiotelis \cite{hiot}, Subramanian, Cen \& Ostriker \cite{subra}).
  Non-radial motions are always present in the structures formed in N-body
  simulations.

  Despite the above assumptions, the results of the model studied
  in this paper are in good agreement with the results of N-body
  simulations. The summary of these results is as follows:
  \begin{enumerate}
  \item Density profiles of haloes are close to the analytical
  models proposed in the literature as good fits to the results of
  N-body simulations. A trend of the inner slope of the density
  profile as an increasing function of the mass of the halo is also
  found, in agreement with recent results of N-body simulations.
  \item Concentration is a decreasing function of virial mass. Its
  values are in agreement with the results of numerical methods.
  \item Massive haloes increase their mass substantially up
  to late times. Growth curves of less massive haloes tend to flatten out
  earlier. The concentrations of less massive haloes evolve more rapidly
  while those of more massive haloes evolve slowly.
  \end{enumerate}
  Taking into account the number of assumptions and approximations
  used to build the model presented in this paper, we can conclude that the agreement with the
  results of N-body simulations is very good. Consequently, this model provides
  a very promising method to deal with the process of structure formation. Further
  improvements to this model could help to understand better the physical picture during
  this  process.

\section{Acknowledgements} I would like to thank the
\emph{Empirikion Foundation} for its financial support.

\bsp

\label{lastpage}

\end{document}